\documentclass[twocolumn,showpacs,preprintnumbers,amsmath,amssymb]{revtex4}
\usepackage{graphicx}
\usepackage{dcolumn}
\usepackage{bm}
\def\bea {\begin{eqnarray}}
\def\eea {\end{eqnarray}}

\def\be {\begin{equation}}
\def\ee {\end{equation}}

\begin{document}
\title{Measuring radial flow of partonic and hadronic phases in 
relativistic heavy ion collision.}

\author{Jajati K Nayak and  Jan-e Alam}

\medskip

\affiliation{Variable Energy Cyclotron Centre, 1/AF, Bidhan Nagar, 
Kolkata - 700064}.

\date{\today}

\begin{abstract}
It has been shown that the thermal photon and the lepton pair spectra
can be used to estimate the radial velocity of different phases of the
matter formed in nuclear collisions at ultra-relativistic energies. We
observe a non-monotonic variation of the  flow velocity with invariant
mass of the lepton pair which is indicative of  two different
thermal dilepton sources at early and late stage of the
dynamically evolving system.  We also show that the study of radial
velocity through electromagnetic probes may shed light on the nature of
the phase transition from hadrons to QGP.
\end{abstract}

\maketitle

\section{Introduction}
The numerical simulation of QCD equation of state (EoS) predicts
that nuclear matter at high density and/or temperature
are composed of quarks and gluons due to asymptotic freedom and
screening of colour charges~\cite{collins,km,evs}.  
Enormous experimental  efforts have been made to produce such a partonic 
state of a matter, called Quark Gluon Plasma (QGP) by colliding nuclei at 
ultra-relativistic energies. Careful
theoretical investigations have been performed to understand 
the existing experimental data~\cite{qm08} and predictions for the
forthcoming experiments~\cite{lhc} have also been made. 

The hot and dense matter formed in the partonic phase after ultra-relativistic 
heavy ion collisions 
expands in space and time due to high internal pressure.
Consequently the system cools and reverts back to hadronic matter from the
partonic phase.
Initially (when the thermal system is just born) 
the entire energy of the system is thermal in nature
and with progress of time some
part of the thermal energy gets converted to the collective (flow) energy.
In other words during the expansion stage 
the total energy of the system is shared by the thermal
as well as collective degrees of freedom. The evolution of 
the collectivity within the system is sensitive to the EoS. 
Therefore, the study of the collectivity in the 
system formed after nuclear collisions will be useful to
shed light on the EoS~\cite{heinz,hung,hirano} of the strongly 
interacting system 
at high temperatures and densities. 

It is well known that the average magnitude of 
radial flow can be extracted from 
the transverse momentum ($p_T)$ spectra of the hadrons.
However, hadrons being strongly interacting objects 
can bring the information of the state of the system
when it is too dilute to support collectivity.
On the other hand electromagnetic (EM) probes {\it i.e.}
photons and dileptons are produced and emitted from
each space time points. Therefore, estimating radial
flow from the EM probes will shed light on the time
evolution of the collectivity in the system.
This is demonstrated by NA60 collaboration
~\cite{NA60} through dilepton measurements in In+In collisions
at SPS energy.
The slope of the transverse mass spectrum of lepton pairs, 
$T_{\mathrm eff}$ of invariant mass 
$M$ can be related to the space-timed averaged quantities like
radial flow velocity $v_{\mathrm r}$ 
and temperature $T_{\mathrm av}$  as
$T_{\mathrm{eff}} \sim T_{\mathrm{av}}+
Mv_{\mathrm r}^2$. $T_{\mathrm eff}$ estimated from 
dilepton spectra~\cite{NA60}
shows a different kind of behaviour 
~\cite{vanhees1,vanhees2,ruppert,renk,dusling,ahns}  
as compared to that from hadronic spectra. 
The effective temperature extracted from transverse mass spectra 
of dileptons increases 
linearly with invariant mass $M$ up to $\rho$-peak and then falls 
(the PHENIX data does not show this trend~\cite{PHENIX}).
In  a recent work~\cite{JKN} we have shown that the ratio ($R_{\mathrm {em}}$)
of the $p_T$ spectra of photons to lepton pairs has 
an advantage over the individual spectra because some of the
uncertainties or model dependence pertaining to the
individual spectra gets canceled in the ratio. Hence the ratio
can be used as an 
efficient tool to understand the state of an expanding system.
In the present work we focus on the extraction of
the radial flow from $R_{\mathrm em}$. 
We also argue that the simultaneous measurements of photons and
dileptons will enable us to estimate the value of $v_{\mathrm r}$
for various invariant mass windows of the lepton pairs. The
$v_{\mathrm r}$ obtained from the analysis of both
the spectra vary with $M$ non-monotonically. 
Such a behaviour may be interpreted as due to the presence 
of two different kinds of thermal sources of lepton pairs 
of the expanding  system.

\par 
The paper is organized as follows. In section II
the ratio of thermal photon and dilepton productions has been discussed. 
In section III the evolution dynamics of the hot fireball system with 
specific initial conditions and EoS is outlined. 
The discussions in sections II and III will be very brief as the details
are available elsewhere~\cite{JKN}. 
The results are presented in section IV. Finally 
section V is devoted to summary and discussions.   
\section{Electromagnetic Probes}
The 
ratio, $R_{\mathrm em}$ of the $p_T$ spectra of thermal photons to dileptons 
can be written as follows~\cite{JKN}:
\begin{equation}
R_{em}= \frac{\frac{d^2N_\gamma}{d^2p_{T}dy}}{\frac{d^2N_\gamma^{\star}}
{d^2p_{T}dy}}
=\frac{\sum_{i}{\int_{i}{\left(\frac{d^2R_\gamma}{d^2p_{T}dy}\right)_id^4x}}}
{\sum_{i}{\int_{i}
{\left(\frac{d^2R_{\gamma^\ast}}{d^2p_{T}dydM^2}\right)_idM^2d^4x.}}
}
\label{rateq}
\end{equation}
The numerator (denominator) is the invariant momentum distribution of the 
thermal photons (lepton pairs). 
In Eq.~\ref{rateq} $p_T$, $y$ and $M$  denote the transverse momentum, 
rapidity and the invariant mass of the lepton pair.
The summation in Eq.~\ref{rateq} runs over 
all phases through which the system passes during the expansion. 
$(d^2R/d^2p_{T}dy)_i$ 
and $(d^2R/d^2p_{T}dydM^2)_i$ are the static rates of
photon and dilepton productions from the phase $i$, which is 
convoluted over the expansion dynamics through the 
space-time integration over $d^4x$.
The integration over $M$ is done by selecting 
invariant mass windows - $M_{\mathrm min}\,\leq\, M\,\leq
M_{\mathrm max}$ appropriately and we define $<M>=(M_{\mathrm min}
+M_{\mathrm max})/2$.

\par
The rate of thermal dilepton production per unit space-time volume 
per unit four momentum volume 
is given by~\cite{mclerran,gale,weldon,alam1}
\begin{equation}
\frac{dR}{d^4p}=\frac{\alpha}{12\pi^4 p^2}L(p^2){\mathrm Im}\Pi_{\mu}^{R\mu}
f_{BE}
\label{eq2}
\end{equation}
where $\alpha$ is the EM coupling constant, ${\mathrm Im}\Pi_{\mu}^{\mu}$ is the
imaginary part of the retarded photon self energy and $f_{BE}(E,T)$ is the
thermal phase space factor for Bosons. 
$L(p^2)$=$(1+\frac{2m^2}{p^2})\sqrt{1-4\frac{m^2}{p^2}}$ arises
from the final state leptonic current involving Dirac spinors of mass $m$.
The real photon production rate can be obtained from the dilepton
emission rate by replacing the product of EM vertex
$\gamma^{\star}$ $\rightarrow$ $ l^{+}l^{-}$, the term involving
final state leptonic current and the square of the (virtual) photon
propagator by the polarization sum for the real photon.
For an expanding system 
$E$ should be  replaced by $u_\mu p^\mu$ where $p^\mu$ and $u^\mu$ are the four 
momentum and the four velocity respectively.

\subsection{Thermal photons}
The photon production rate has been evaluated by various
authors~\cite{photons} using Hard Thermal Loops~\cite{braaten} approximations.
The complete calculation of emission rate of photons from QGP to order
O$(\alpha,\alpha_s)$ has been done 
by resuming ladder diagrams in the effective theory~\cite{arnold}.
This rate of production 
has been considered in the present work. A set of  hadronic reactions
with all isospin combinations have been considered for the production of
photons~\cite{npa1,npa2,turbide} from hadronic matter. 
The effect of hadronic dipole
form factors has been taken into account in the present
work.
We have checked that the high $p_T$ ($\sim 2-3$ GeV) part of the 
thermal photon spectra is dominated
by the contributions from QGP phase for large initial temperature.

\subsection{Thermal dileptons}
The lowest order process producing lepton pairs is $q$ and $\bar{q}$
annihilation. For a finite temperature QCD plasma, the correction 
of order $\alpha_s\alpha^2$ to
the lowest order rate of dilepton production has been calculated in
~\cite{altherr,thoma}, which is considered in the present work.
For the low $M$ dilepton production from the hadronic phase we consider the
decay of light vector mesons $\rho, \omega$ and $\phi$ as considered
in ~\cite{JKN}. The continuum part of the vector mesons spectral functions have
been included in the present work~\cite{annals,shuryak}.

It is well known that the contributions from the  QGP phase dominates 
the $M$ spectra of the lepton pairs below $\rho$-peak 
and above the $\phi$-peak if no thermal effects of the 
spectral functions 
of the vector mesons~(see \cite{BR,RW,annals} for review) are considered.
 

\section{Evolution dynamics}
In the collision of two energetic heavy ions a large amount
of energy is dumped into a small volume. 
The space-time evolution of the matter has
been studied using ideal relativistic hydrodynamics 
~\cite{von} with longitudinal boost invariance ~\cite{bjorken}
and cylindrical symmetry. 
The initial energy density ($\epsilon(\tau_i,r)$) and radial velocity
($v(\tau_i,r)$)  profiles are same as our earlier studies~\cite{JKN}. 
The value of
transition temperature ($T_c$) is  taken as 192 MeV as obtained
in lattice QCD calculations ~\cite{cheng}. Although a much lower
value of $T_c$ is also predicted in~\cite{fodor}.
However, we have found that the dependence of $R_{\mathrm em}$ 
on $T_c$ is  weak. 
In a first order phase transition scenario we use Bag EoS for the QGP 
phase and for the hadronic phase all the resonances with mass $\le 2.5$ GeV 
have been considered~\cite{bm}.  

In the present work we have considered the initial and freeze out conditions 
that 
reproduced the hadrons~\cite{npa2002}, the photon ~\cite{JA07} and dilepton 
spectra 
for RHIC energy \cite{we2}. The values of initial thermalization time,
$\tau_i=0.2$  fm/c, initial temperature $T_i=400$ MeV and the freeze-out 
temperature $T_F=120$ MeV have been taken as the input to the calculation.
For LHC we have taken $T_i=700$ MeV, $\tau_i=0.08$ fm/c which gives the 
hadron multiplicity $dN/dy=2100$~\cite{lhc}.

\begin{figure}
\begin{center}
\includegraphics[scale=0.43]{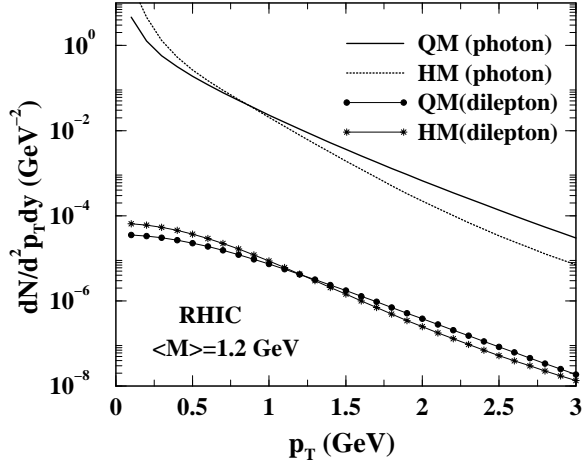}
\caption{The $p_T$ spectra of photons and dileptons from hadronic and 
quark matter at RHIC energy. The dilepton spectra is obtained by 
doing $M$ integration from $M=$1.0 GeV to 1.4 GeV}.
\label{fig:fig1}
\end{center}
\end{figure} 
\section{Results and Discussion}
In Fig.~\ref{fig:fig1} the photon and dilepton spectra 
have been displayed for RHIC conditions. Results indicate that 
the photon spectra from QGP dominates over its hadronic
counterpart for $p_T > 1.5$ GeV. The dilepton from QGP
and hadrons are comparable in magnitude for entire range of
$p_T$ for $M\sim 1.2$ GeV (this is because of the inclusion of
the continuum of the vector meson spectral functions~\cite{shuryak,annals}
without the continuum the quark matter part dominates). 
However, for $M\sim 0.75$ GeV
the dileptons from the hadronic matter are overwhelmingly large 
compared to quark matter contributions (not shown in the figure).  
Therefore, an appropriate
selection of $p_T$ and $M$ will be very useful to characterize
a particular phase of the system.

Now we consider the variation of the ratio, $R_{em}$ as a function of $p_T$ 
for different invariant mass
windows. The results  are shown in the Fig.~\ref{fig:fig2} and 
Fig.~\ref{fig:fig3}
below. 
The variation of $R_{\mathrm em}$ with respect to $p_T$ can be 
parametrized as follows:
\bea
{R}_{em} & \equiv & A_{3} [\frac{m_T}{p_T}]^{B_{3}}
exp[C_{3}(m_T-p_T)]\nonumber\\ 
\label{eqflow1}
\eea
where $A_3$, $B_3$ and $C_3$ are constants and $M_T$, the transverse mass 
of the lepton pair 
is defined as, $M_T=\sqrt{p_T^2+M^2}$. 
It is observed that the ratio decreases sharply 
and reaches a plateau beyond $p_T$ $>$ 1.5 GeV. 
This behaviour of $R_{\mathrm em}$
as a function of $p_T$ can be understood as follows: 
(i) for $p_T>>M\,\, M_T\sim 
p_T$ and consequently $R_{\mathrm em}\sim A_3$ giving rise to a plateau
at large $p_T$. The height of the plateau is sensitive to the initial 
temperature of the system~\cite{JKN}.
(ii) For $p_T<M$ $R_{\mathrm em}\sim exp(-p_T/T_{\mathrm eff})/
p_T^{\mathrm {B_3}}$
indicating a decrease of the ratio with $p_T$ (at low $p_T$) as observed 
in the     
Fig.~\ref{fig:fig2} and Fig.~\ref{fig:fig3}.  

To indicate the effect of the radial flow velocity, $v_{\mathrm r}$  
we have evaluated 
the $R_{\mathrm em}$ with and without radial flow 
(see Fig.~\ref{fig:fig2} and Fig.~\ref{fig:fig3}).
In case of vanishing radial flow the ratio
can be parametrized as follows:
\bea
{R^1}_{em} & \equiv & A_{1} [\frac{m_T}{p_T}]^{B_{1}}
exp[C_{1}(m_T-p_T)]\nonumber\\  
\label{eqflow4}
\eea
Here $C_1$ contains the information of the average temperature, $T_{av}$ 
of the system. 

In case of vanishing radial flow velocity the 
inverse slope of the photon and dilepton spectra represent 
the average temperature, $T_{\mathrm av}$ of the system. However, in case of 
non-zero radial flow the inverse slope contains the effect of 
average temperature as well as that of $v_{\mathrm r}$. 
Therefore, the difference in the slopes of the two cases 
will enable us to estimate the amount of collectivity  in the
system. 

As mentioned before
for large initial temperature transverse momentum distribution of
photons from QGP dominates over its hadronic counterpart for 
$p_T\geq 1.5 $ GeV. However,
in case of dileptons one has to select both the $M$ and the
$p_T$ windows to observe QGP. For example the thermal
dileptons from hadrons dominate over those from QGP
for $M\sim 0.75$ GeV. 
Therefore, for estimating 
the radial velocity in the hadronic phase we chose $p_T\sim 0.5$ 
GeV and $M\sim 0.75$ GeV for demonstrative purpose.
Similarly a $p_T$ and $M$ windows may 
be selected where contributions from QGP dominates. 

\begin{figure}
\begin{center}
\includegraphics[scale=0.43]{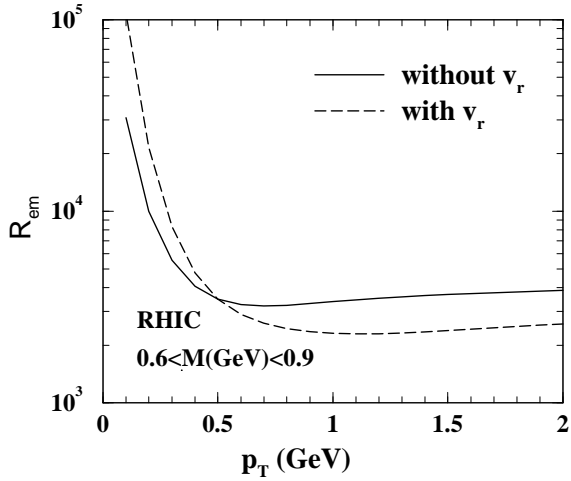}
\caption{$R_{\mathrm em}$ as a function of $p_T$ with and without 
radial flow for 
invariant mass $0.6<M(GeV)<0.9$. 
The spectra with radial flow is normalized to the one 
without radial flow at $p_T=0.5$ GeV 
}
\label{fig:fig2}
\end{center}
\end{figure} 
\begin{figure}
\begin{center}
\includegraphics[scale=0.43]{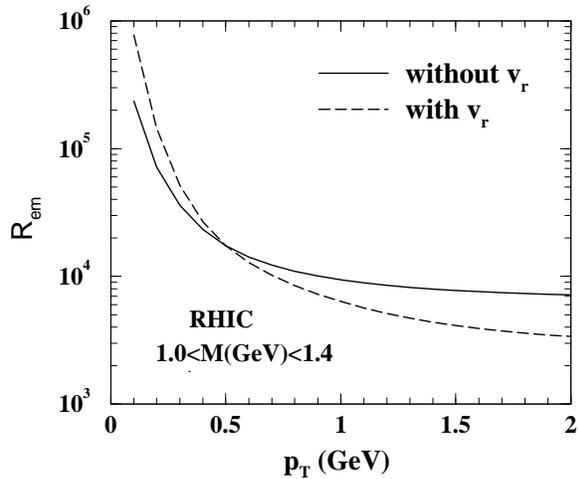}
\caption{$R_{\mathrm em}$ as a function of $p_T$ 
as in the above figure for
invariant mass $1.0<M(GeV)<1.4$. 
}
\label{fig:fig3}
\end{center}
\end{figure} 
\par
\begin{figure}
\begin{center}
\includegraphics[scale=0.43]{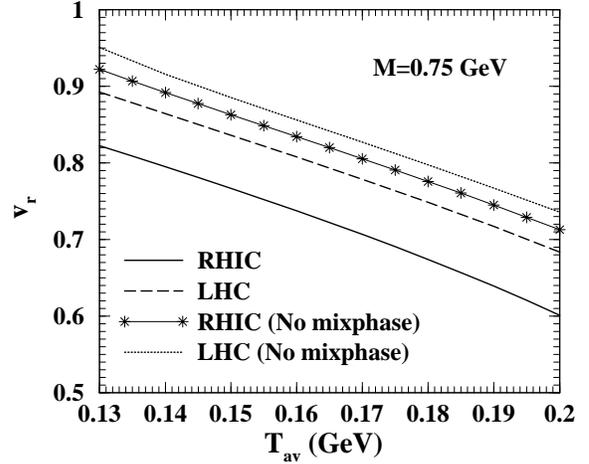}
\caption{Variation of $v_{\mathrm r}$ with $T_{\mathrm av}$ for 
$M=0.75$ GeV and $p_T=0.5$ GeV. The solid (dashed) line indicate the results 
for RHIC (LHC) 
for EoS with
first order phase transition. The line with 
asterisk (dotted line) stands for RHIC (LHC)for an EoS which excludes the mixed phase.   
} 
\label{fig:fig4}
\end{center}
\end{figure}
\begin{figure}
\begin{center}
\includegraphics[scale=0.43]{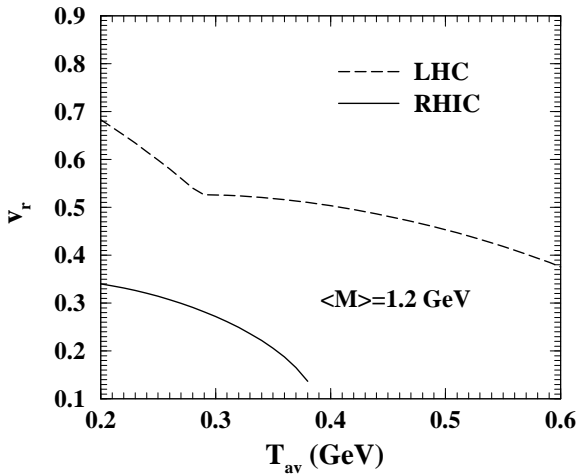}
\caption{
Variation of $v_{\mathrm r}$ with $T_{\mathrm av}$ for $M=1.2$ GeV
and $p_T=0.5$ GeV at RHIC and LHC energies for EoS with a first order phase
transition.} 
\label{fig:fig5}
\end{center}
\end{figure}

The exponential slope of the ratio ($C_3$)  can be related to the 
individual slopes of photons ($T_{\mathrm eff1}^{-1}$) and dileptons 
($T_{\mathrm eff2}^{-1}$) as follows:
\begin{equation}
C_3 \times {(m_T-p_T)}=\frac{m_T}{T_{eff2}}-\frac{p_T}{T_{eff1}} \nonumber\\ 
\end{equation}
writing the effective (blue shifted) temperatures of the photon spectra and 
dilepton spectra as

\begin{equation}
T_{eff1}=T_{av}\sqrt{\frac{(1+v_r)}{(1-v_r)}}, \nonumber\\
\end{equation}
\begin{equation}
T_{eff2}=T_{av}+M{v_r}^2
\label{eqntemp}
\end{equation}
we obtain,
\begin{equation}
C_3 \times {(m_T-p_T)}=\frac{m_T}{T_{av}+M{v_r}^2}-
\frac{p_T}{T_{av}\sqrt{(1+v_r)/(1-v_r)}} \nonumber\\
\end{equation}
Further simplification leads to   
\begin{equation}
aT_{{av}}^2+bT_{av}+c=0
\label{eqnflow}
\end{equation} 
where $a$, $b$ and $c$ are functions of $v_{\mathrm r}$.
Solving Eq.~\ref{eqnflow}  for a given $C_3$, $M$ and $p_T$
we obtain $v_{\mathrm r}$ as a function of the average temperature.
The results are displayed in Figs.~\ref{fig:fig4} and ~\ref{fig:fig5}
for initial conditions of RHIC and LHC energies
for invariant mass and $p_T$ windows indicated.  
The contributions in the $M$ and $p_T$ windows shown in Fig.~\ref{fig:fig4}
are dominated by the hadronic phase {\it i.e.} from temperature range
$T_c\sim 192 $ MeV to $T_F\sim 120$ MeV. The radial velocity 
increases sharply with decrease in $T_{\mathrm av}$ in the hadronic phase. 

We have evaluated $v_{\mathrm r}$ with a (continuous) EoS where the mixed phase
does not appear. In this case the $v_{\mathrm r}$ is larger than
the one obtained for a strong first order phase transition 
(~Fig.~\ref{fig:fig4}).  
Indicating that the presence of the mixed phase (of hadrons and QGP)
characterized by zero sound velocity slowed down the  expansion
of the system, resulting in a lower radial flow. Therefore, extraction of 
$v_{\mathrm r}$ from experimental data will be useful to understand the 
nature of the transition.

In Fig.~\ref{fig:fig5} the radial velocity is displayed 
for (average) temperature range which is dominated by QGP phase. 
The results indicate a  moderate $v_r$ for RHIC but a large $v_r$  is 
achieved even in the QGP phase for LHC energies, in fact a fast increase 
in $v_{\mathrm r}$ is observed for $T_{\mathrm av}$ close to the transition 
temperature in case of LHC. The $v_{\mathrm r}$ for LHC is much
larger than RHIC because of the longer life time and larger 
internal pressure of the partonic phase in LHC than RHIC. 

\begin{figure}
\begin{center}
\includegraphics[scale=0.4]{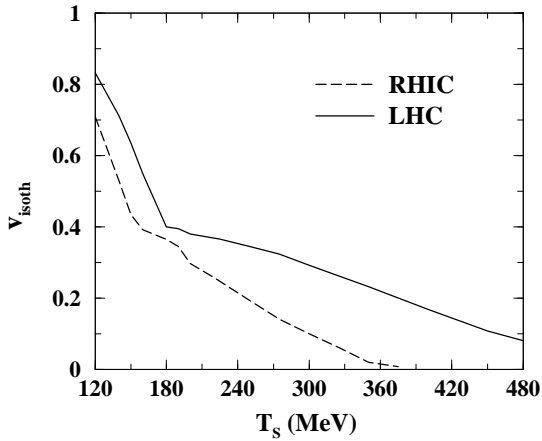}
\caption{Variation of  average radial velocity of the fluid
on the constant temperature surface. 
}
\label{fig:fig6}
\end{center}
\end{figure}

\begin{figure}
\begin{center}
\includegraphics[scale=0.4]{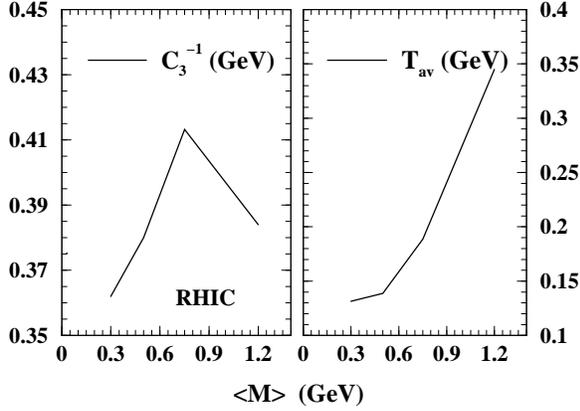}
\caption{Left panel: The variation of the slope, $C_3$ with invariant mass 
obtained from the $p_T$ spectra of ratio for RHIC energy is displayed in 
the left panel of the curve. Right panel: the variation of average 
temperature of the system. The left (right) vertical label is for left (right) 
panel of the curve} 
\label{fig:fig7}
\end{center}
\end{figure}

\begin{figure}
\begin{center}
\includegraphics[scale=0.4]{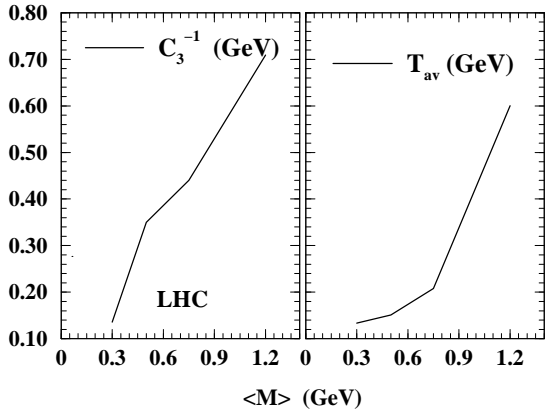}
\caption{The variation of the slope, $C_3$ with invariant mass obtained 
from the $p_T$ spectra of ratio for LHC. Please note that the scales
in the left and the right panels are same.} 
\label{fig:fig8}
\end{center}
\end{figure}

In a first order phase transition scenario the QGP formed in heavy ion
collisions return back to hadrons through a first order phase transition.
The temperature changes continuously from $T_i$ to $T_F$. We estimate
the average values of the radial velocity ($v_{\mathrm isoth}$) on the 
constant temperature surfaces determined by the conditions: $T(r,\tau)=T_S$, 
for various values $T_S$. The variation of  ($v_{\mathrm isoth}$)  with
$T_S$ is depicted in Fig.~\ref{fig:fig6} both for RHIC and LHC 
energies. $v_{\mathrm isoth}$ for LHC is larger than RHIC because 
of higher initial temperature and hence internal pressure.
In contrast to the results
shown in Figs.~\ref{fig:fig4} and ~\ref{fig:fig5},  
the variation of $v_{\mathrm isoth}$ with $T_S$ is not measurable
as it does not depend on the kinematic variables, $p_T$ and $M$.
The expansion is slower in the hadronic phase because of the softer
EoS as compared to the QGP phase. For given $T_c$ and $T_F$ the
life time of the hadronic phase is larger for softer EoS - allowing
the system to develop large radial flow  as evident from the
results depicted in Fig.~\ref{fig:fig6} for the low temperature 
part. 
The effective temperature extracted from the ratio is displayed in  
Fig.~\ref{fig:fig7} as a function of $M$ for RHIC energy. 
$T_{\mathrm eff}$ increases with $M$ up to the $\rho$-peak
and then decreases beyond $\rho$ mass. The reduction of $T_{\mathrm eff}$
beyond $\rho$ is indicating
the dominance of the radiation from the high temperature phase 
in the high $M$ region. 
For LHC, however,  no clear reduction of $T_{\mathrm eff}$ beyond $\rho$ - peak
is observed (Fig.~\ref{fig:fig8}). 
At LHC the average temperature  and the flow velocity in the 
early phase (from where large $M$ pairs originate) are large 
(see Fig.~\ref{fig:fig4}).
Hence the combination of both large $v_r$ and
large $T_{av}$ does not allow $T_{\mathrm eff}$ 
to fall above the $\rho$-peak.
The dependence of individual spectra on $T_F$ is 
quite strong, however, we have observed that
the slope of the ratio is insensitive to $T_F$ and also to $T_c$.
The slope of the ratio does not change when the 
parameters like $T_F$ changes  from $0.120$ GeV to 
$0.150$ GeV and $T_c$ from $0.192$ GeV to $0.175$ GeV. 

\begin{figure}
\begin{center}
\includegraphics[scale=0.42]{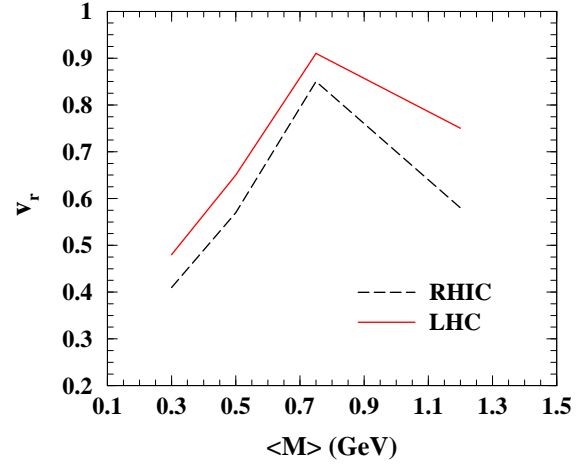}
\caption{Radial velocity as a function of $M$ for RHIC and LHC energies.}
\label{fig:fig9}
\end{center}
\end{figure}

Eliminating $T_{\mathrm av}$ from Eq.~\ref{eqntemp} and taking the values
of $T_{\mathrm eff1}$ and $T_{\mathrm eff2}$ from photon and dilepton
spectra one can obtain the variation of $v_r$ as a function of $M$.
The results are shown in Fig.~\ref{fig:fig9} for RHIC and LHC energies.
A non-monotonic behaviour of $v_r$ with $M$ is observed.
A similar non-monotonic behaviour is observed in the elliptic
flow ($v_2$) of photons as a function of transverse momentum~\cite{rupa,liu}.
Comparison of dilepton production from QGP and hadronic sources~\cite{JKN}
indicate that in the low $M (<m_\rho)$ and high $M (>m_\phi)$   region
the emission rate from QGP dominates over its hadronic counter part if the
medium effects on the vector meson spectral functions are neglected. 
In other words, for a dynamically evolving system 
the low and high $M$ pairs are emitted from early 
QGP phase, whereas lepton pairs with $M$ around $\rho$ - mass are emitted
from the late hadronic phase. Therefore, low and high $M$ domains 
represent early time where $v_{\mathrm r}$ is low and the $M\sim m_\rho$ 
domain represent late time - where $v_{\mathrm r}$ is large - giving rise
to the observed variation in Fig.~\ref{fig:fig9}- indicative of a two different
kinds of source in early and late times of the evolving system.
For $M \sim 1.2$  GeV 
the flow velocity is not very small since this window is  
populated by both hadronic and partonic contributions almost equally.  
Again at LHC energy the partonic phase life time is more which favors the 
development of larger flow compared to RHIC energy. 
It is important to note at this point that for LHC, although the slope $C_3$
does not show a clear non-monotonic behaviour with $M$, $v_r$ does so.
Because as described, before the slope $C_3$ depends not only on $v_r$
 but also on $T_{av}$ and both are large in the partonic phase at LHC.

\begin{figure}
\begin{center}
\includegraphics[scale=0.4]{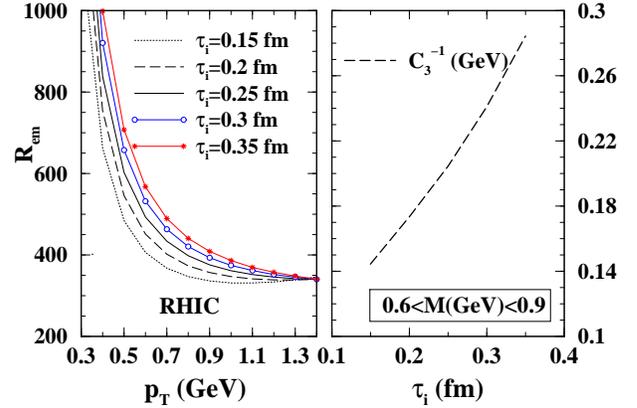}
\caption{Left panel: Ratio of the $p_T$ spectra for different 
initial thermalization 
time $\tau_i$ with all other parameters kept same. Right panel: variation of the
effective slope $C_3$ as a function of initial thermalization time, $\tau_i$.
The left (right) vertical label is for left (right) 
panel of the curve} 
\label{fig:fig10}
\end{center}
\end{figure}

The two time scales - the life time of the 
partonic phase ($\tau_{\mathrm QGP}$)  and the time an inward
moving rarefaction wave takes to
hit the centre of the cylindrical geometry decide whether 
radial flow play any important role in the partonic phase or not. 
The later time scale
is defined as $\tau_{\mathrm rw}\sim R/c_s$ where $R$ is the transverse
size of the system and $c_s$ is the velocity of sound.
If $\tau_{\mathrm QGP}\sim\tau_{\mathrm rw}$ then $v_{\mathrm r}$ will
be large in the partonic phase.  
Therefore, an increase in $\tau_i$ ($\tau_{\mathrm QGP}\propto \tau_i$) 
will increase the radial
flow in the partonic phase if the initial and the critical temperatures
are kept fixed. However, an increase in $\tau_i$ from 
$\tau_1$ to $\tau_2$ produces same flow if the $T_i$ decreases by a factor 
$(\tau_2 /\tau_1 )^{1/3}$. For a fixed $T_i$ an increase in $\tau_i$
will increase the effective slope as evident from the right panel of Fig.~\ref{fig:fig10}.
Therefore, the slope of the ratio may be used effectively to estimate the value of initial
thermalization time.

\section{Summary and Discussions}
It has been shown that the $p_T$ distribution of
thermal photons and lepton pair spectra 
may be used simultaneously to estimate the magnitude of  the radial velocity 
of different phases of the matter formed 
in nuclear collisions at ultra-relativistic energies. 
Judicious choices of the kinematic variables {\it e.g.} the invariant mass
and the transverse momentum windows may be selected  to estimate the
flow velocity in the partonic and hadronic phases of the evolving 
matter.  It has been observed that for RHIC and LHC energies the flow 
velocity increases with invariant mass up to the $\rho$ peak beyond  
which it decreases. 
The $T_{\mathrm eff}$ may not decrease with mass beyond $\rho$  peak if the 
average temperature and the flow velocity are large in the 
partonic phase as in case of LHC energy.   
By doing a  simple analysis of photon and dilepton 
spectra we have extracted the radial flow 
velocity for various invariant mass windows. $v_{\mathrm r}$ varies 
with $M$ non-monotonically. We argue that such a variation indicates
the presence of two different types of thermal sources of lepton pairs. 

{\bf Acknowledgment:} JA is  supported by DAE-BRNS
project Sanction No.  2005/21/5-BRNS/2455.

\end{document}